\documentclass[aps,prd,onecolumn,groupedaddress,showpacs,nofootinbib,amssymb]{revtex4}
\usepackage[dvips]{graphicx}
\usepackage{amssymb}
\usepackage{amsmath}
\usepackage{graphicx}
\usepackage{amsfonts}
\usepackage{bm}
\usepackage{amsmath,amssymb,theorem}
\begin{document}

\title{Effects of Spatial Curvature on the $f(R)$ Gravity Phase Space: no Inflationary Attractor?}
\author{S.~D. Odintsov,$^{1,2}$\,\thanks{odintsov@ieec.uab.es}
V.K. Oikonomou,$^{3,4,5}$\,\thanks{v.k.oikonomou1979@gmail.com}}

\affiliation{$^{1)}$ ICREA, Passeig Luis Companys, 23, 08010 Barcelona, Spain\\
$^{2)}$ Institute of Space Sciences (IEEC-CSIC) C. Can Magrans s/n,
08193 Barcelona, Spain\\
$^{3)}$ Department of Physics, Aristotle University of
Thessaloniki, Thessaloniki 54124,
Greece\\
$^{4)}$ Laboratory for Theoretical Cosmology, Tomsk State
University of Control Systems
and Radioelectronics, 634050 Tomsk, Russia (TUSUR)\\
$^{5)}$ Tomsk State Pedagogical University, 634061 Tomsk, Russia\\
}

\tolerance=5000

\begin{abstract}
In this paper we study the effects of spatial curvature of the
metric on the phase space of vacuum $f(R)$ gravity. Particularly,
we appropriately choose the variables of the dynamical system, in
order for this to be autonomous, and we study the phase space of
the resulting theory, focusing on de Sitter, matter and radiation
domination fixed points. Our analysis indicates that the effect of
spatial curvature on the phase space is radical, since it
destabilizes all the stable de Sitter vacua of the flat spacetime
vacuum $f(R)$ gravity phase space, making the phase space having
non-trivial unstable submanifolds. This instability occurs
regardless if the spacetime has elliptic or hyperbolic spatial
sections, and it is also robust towards the choice of initial
conditions. We investigate the source of the instability in the
system, and also we discuss the stability of the matter and
radiation domination vacua, which, as we demonstrate, are also
highly unstable. Our results for de Sitter attractors indicate
that the stable de Sitter attractors of the vacuum $f(R)$ gravity
theory for a flat Universe, are destabilized by the presence of
curvature, and this shows that inflation for vacuum $f(R)$ gravity
in non-flat spacetime is problematic, at least at the phase space
level. This result holds true for both elliptic and hyperbolic
spacetimes.
\end{abstract}


\maketitle

\section{Introduction}

Among the various forms of modified gravity, $f(R)$ gravity
provides the simplest and most elegant to our opinion, description
of several evolutionary aspects of our Universe
\cite{reviews1,reviews2,reviews3,reviews4,reviews5,reviews6}.
Indeed, it is possible to describe two of the most mysterious
acceleration eras of our Universe, namely the inflationary and the
late-time acceleration era, in a unified framework
\cite{Nojiri:2003ft,Nojiri:2006gh}, and several works in the
literature investigate the astrophysical and cosmological
implications of $f(R)$ gravity, for an important stream of viable
$f(R)$ gravity models, see for example
\cite{Nojiri:2003ft,importantpapers1,importantpapers2,importantpapers4,importantpapers5,importantpapers8,importantpapers9,importantpapers10,importantpapers11,importantpapers11a,
importantpapers12,importantpapers13,importantpapers14,importantpapers14aaa,importantpapers15,importantpapers17,importantpapers20}.
In view of the importance of $f(R)$ gravity in modern cosmology
applications, it is vital to know the general behavior of $f(R)$
gravity in a model independent way, and the only formal and
mathematically rigid way to achieve this is to study the phase
space of $f(R)$ gravity. The correct approach for doing so, is to
construct an autonomous dynamical system for the simplest $f(R)$
gravity action, namely that of vacuum $f(R)$ gravity. Actually the
dynamical system approach is frequently adopted in the literature
for studying cosmological systems with cosmological applications
in the study of inflation and dark energy for modified gravity and
scalar-tensor theories and cosmological fluids, see for example
\cite{ellis,Bahamonde:2017ize,Boehmer:2014vea,Carloni:2018yoz,Bohmer:2010re,Goheer:2007wu,Leon:2014yua,Leon:2010pu,deSouza:2007zpn,Giacomini:2017yuk,Kofinas:2014aka,Leon:2012mt,Gonzalez:2006cj,Alho:2016gzi,Biswas:2015cva,Muller:2014qja,Mirza:2014nfa,Rippl:1995bg,Ivanov:2011vy,Khurshudyan:2016qox,Boko:2016mwr,Odintsov:2017icc,Granda:2017dlx,Landim:2016gpz,Landim:2015uda,Landim:2016dxh,Bari:2018edl,Chakraborty:2018bxh,Ganiou:2018dta,Shah:2018qkh,Odintsov:2018uaw,Odintsov:2018awm,Oikonomou:2017ppp,Odintsov:2017tbc,Kleidis:2018cdx,Kleidis:2018fdu,Odintsov:2015wwp,Dutta:2017fjw,Gosenca:2015qha,vandenHoogen:1999qq,Perez:2013zya,Goliath:1998na}.
The autonomous dynamical system approach was adopted in several
previous works on $f(R)$ gravity \cite{Odintsov:2017tbc} and other
modified gravities \cite{Oikonomou:2017ppp,Kleidis:2018cdx}, for
the study of de Sitter eras and also for matter domination eras.
To our opinion, the autonomous dynamical system approach is the
most correct approach in studying dynamical systems, since if a
dynamical system is non-autonomous, then the linearization
techniques, which hold true for autonomous dynamical systems, like
the Hartman-Grobman theorem, may lead to mathematical
inconsistencies and wrong results. For example
\cite{dynsystemsbook}, the dynamical system $\dot{x}=-x+t$ has an
explicit solution $x(t)=t-1+e^{-t}(x_0+1)$, hence all the
solutions corresponding to various initial conditions,
asymptotically approach $t-1$. However, if one applies the
linearization theorems, one finds that the fixed point is time
dependent and it is $x=t$, which is not even a solution to the
initial dynamical system. In our previous work on vacuum $f(R)$
gravity autonomous dynamical system \cite{Odintsov:2017tbc}, the
whole study was performed by assuming a spatially flat
Friedmann-Robertson-Walker (FRW) geometric background.  However
due to the fact that there exist several studies which indicate
that a spatially curved and actually closed $\Lambda$CDM model,
provide better fit to the low multipole CMB data
\cite{Ooba:2018dzf,Ooba:2017lng}\footnote{However high multipole
data support the flat $\Lambda$CDM.}, in this work, we will
concretely examine the effects of spatial curvature on the phase
space corresponding to the autonomous dynamical system of vacuum
$f(R)$ gravity. In the literature there exist many works focusing
on the effects of spatial curvature on the phase space, see for
example
\cite{Gosenca:2015qha,vandenHoogen:1999qq,Perez:2013zya,Goliath:1998na,ellis}.
The focus will be on the existence and stability of de Sitter
vacua and also for other possible attractor solutions, like matter
and radiation domination attractors. Our study will be focused on
the first $60-70$ $e$-foldings, which is a characteristic time
interval for inflationary attractors. In most of the cases, the
presence of the spatial curvature will make the phase space quite
more complicated, so we shall rely to numerical analysis in order
to examine the behavior of the trajectories in the phase space. As
in the vacuum $f(R)$ gravity in flat FRW spacetime, the
time-dependence in the dynamical system is contained in the
parameter $m=-\frac{\ddot{H}}{H^3}$, so by assuming that this
parameter is constant, we examine a subclass of the phase space
which can actually yield such a value for $m$, and we seek for
stable regions or fixed points for this subclass of $f(R)$ gravity
cosmologies. In contrast to the flat vacuum $f(R)$ gravity
autonomous dynamical system, the spatially curved autonomous
$f(R)$ gravity dynamical system has no stable de Sitter
attractors, nor has stable attractors of matter and radiation
domination type. As we will demonstrate in our detailed analysis,
spatial curvature induces strong instabilities in the phase space.
This result clearly indicates that the stable inflationary
attractors of the flat vacuum $f(R)$ gravity cease to be stable in
the non-flat vacuum $f(R)$ gravity, regardless if the curvature is
positive (elliptic spatial sections) or negative (hyperbolic
spatial curvatures). This result can be interpreted that inflation
for $f(R)$ gravity in non-flat spacetimes seems to be problematic.

This paper is organized as follows: In section I we review the
properties of the flat space autonomous dynamical system of vacuum
$f(R)$ gravity, emphasizing on the de Sitter fixed points and
their stability. In section II, we construct the non-spatially
flat autonomous dynamical system of vacuum $f(R)$ gravity, and we
investigate the existence of de Sitter, matter domination and
radiation domination fixed points. We examine the stability of the
fixed points and we compare the non-flat dynamical system with the
flat dynamical system. Also we solve numerically the dynamical
system and investigate the behavior of the trajectories in the
phase space as a function of the $e$-foldings number $N$,
emphasizing in values in the range $N=[0,60]$. Finally the
conclusions follow in the end of the paper.

\section{Autonomous Dynamical System of Vacuum $f(R)$ Gravity in Flat Spacetime: An Overview}

Before we get to the core of this paper, it is worth highlighting
the most important outcomes of the flat autonomous $f(R)$ gravity
dynamical system study, and for details we refer the reader to
Ref. \cite{Odintsov:2017tbc}. The general vacuum $f(R)$ gravity
action is,
\begin{equation}\label{action}
\mathcal{S}=\frac{1}{2\kappa^2}\int \mathrm{d}^4x\sqrt{-g}f(R)\, ,
\end{equation}
with $\kappa^2=8\pi G=\frac{1}{M_p^2}$ and in addition $M_p$
stands for the Planck mass scale. In this work we shall adopt the
metric formalism, so by  varying the action (\ref{action}) with
respect to the metric $g_{\mu \nu}$, we get the equations for
motion for vacuum $f(R)$ gravity,
\begin{equation}\label{eqnmotion}
F(R)R_{\mu \nu}(g)-\frac{1}{2}f(R)g_{\mu
\nu}-\nabla_{\mu}\nabla_{\nu}F(R)+g_{\mu \nu}\square F(R)=0\, ,
\end{equation}
which can be rewritten,
\begin{align}\label{modifiedeinsteineqns}
R_{\mu \nu}-\frac{1}{2}Rg_{\mu
\nu}=\Big{(}\frac{1}{F(R)}\Big{(}\frac{f(R)-RF(R)}{2}g_{\mu
\nu}+\nabla_{\mu}\nabla_{\nu}F(R)-g_{\mu \nu}\square
F(R)\Big{)}\Big{)}\, ,
\end{align}
where the prime indicates differentiation with respect to $R$, the
Ricci scalar. The flat FRW metric has the following line element,
\begin{equation}\label{frw}
ds^2 = - dt^2 + a(t)^2 \sum_{i=1,2,3} \left(dx^i\right)^2\, ,
\end{equation}
where $a(t)$ is the scale factor, and the corresponding Ricci
scalar is,
\begin{equation}\label{ricciscalaranalytic}
R=6\left (\dot{H}+2H^2 \right )\, ,
\end{equation}
where $H=\frac{\dot{a}}{a}$ is the Hubble rate. By applying the
FRW metric in (\ref{modifiedeinsteineqns}), the equations of
motion take the following form,
\begin{align}
\label{JGRG15} 0 =& -\frac{f(R)}{2} + 3\left(H^2 + \dot H\right)
F(R) - 18 \left( 4H^2 \dot H + H \ddot H\right) F'(R)\, ,\\
\label{Cr4b} 0 =& \frac{f(R)}{2} - \left(\dot H + 3H^2\right)F(R) +
6 \left( 8H^2 \dot H + 4 {\dot H}^2 + 6 H \ddot H + \dddot H\right)
F'(R) + 36\left( 4H\dot H + \ddot H\right)^2 F'(R) \, ,
\end{align}
with $F(R)=\frac{\partial f}{\partial R}$, $F'(R)=\frac{\partial
F}{\partial R}$, and $F''(R)=\frac{\partial^2 F}{\partial R^2}$.
We can form an autonomous dynamical system for the vacuum $f(R)$
gravity, by using the following dimensionless variables,
\begin{equation}\label{variablesslowdown}
x_1=-\frac{\dot{F}(R)}{F(R)H},\,\,\,x_2=-\frac{f(R)}{6F(R)H^2},\,\,\,x_3=
\frac{R}{6H^2}\, ,
\end{equation}
so by using the $e$-foldings number instead of the cosmic time, as
a dynamical variable that quantifies the evolution, after some
algebra we obtain the following dynamical system for the flat
vacuum $f(R)$ gravity,
\begin{align}\label{dynamicalsystemmain}
& \frac{\mathrm{d}x_1}{\mathrm{d}N}=-4+3x_1+2x_3-x_1x_3+x_1^2\, ,
\\ \notag &
\frac{\mathrm{d}x_2}{\mathrm{d}N}=8+m-4x_3+x_2x_1-2x_2x_3+4x_2 \, ,\\
\notag & \frac{\mathrm{d}x_3}{\mathrm{d}N}=-8-m+8x_3-2x_3^2 \, ,
\end{align}
where we used Eqs. (\ref{variablesslowdown}) and (\ref{JGRG15}),
and in addition, the following differentiation rule,
\begin{equation}\label{specialderivative}
\frac{\mathrm{d}}{\mathrm{d}N}=\frac{1}{H}\frac{\mathrm{d}}{\mathrm{d}t}\,
.
\end{equation}
Moreover, the parameter $m$ appearing in Eq.
(\ref{dynamicalsystemmain}) is equal to,
\begin{equation}\label{parameterm}
m=-\frac{\ddot{H}}{H^3}\, ,
\end{equation}
so the only time dependence (or equivalently $N$-dependence) is
contained in the parameter $m$. Therefore, in the cases that
$m=$constant, the phase space solutions are narrowed down and the
dynamical system (\ref{dynamicalsystemmain}) becomes autonomous.
It is conceivable that when we specify the parameter $m$ to be
constant, in some sense we seek for a class of cosmological
solutions of a specific form, hence by studying the dynamical
system (\ref{dynamicalsystemmain}) for the given value of $m$,
clearly reveals whether the vacuum $f(R)$ gravity theory can
realize such a class of solutions, and also indicates the behavior
of the trajectories in the phase space, or equivalently, such a
study reveals the existence and stability of cosmological fixed
points in the vacuum $f(R)$ gravity phase space. In the flat
vacuum $f(R)$ gravity case, the equation of state (EoS) parameter
$w_{eff}$, which is in general defined to be,
\begin{equation}\label{weffoneeqn}
w_{eff}=-1-\frac{2\dot{H}}{3H^2}\, ,
\end{equation}
when expressed in terms of the variables
(\ref{variablesslowdown}), can be written as follows,
\begin{equation}\label{eos1}
w_{eff}=-\frac{1}{3} (2 x_3-1)\, .
\end{equation}
In Ref. \cite{Odintsov:2017tbc} we focused our analysis in the
cases $m=0$ and $m=-\frac{9}{2}$, which correspond in the de
Sitter and matter domination cases. Note that the quasi-de Sitter
evolution $a(t)=e^{H_0 t-H_i t^2}$, along with the exact de Sitter
evolution $a(t)=e^{\Lambda t}$, can yield $m=0$, but these are not
the only cosmologies that may yield $m=0$, for example the
symmetric bounce may also yield $m=0$. The nature of the
attractors of the phase space for $m=0$ will be determined solely
by the value of the effective EoS parameter $w_{eff}$ at the
resulting fixed point. At this point we need to further clarify
the case $m=$const, focusing for example on the case $m=0$. By
solving the differential equation (\ref{parameterm}) for $m=0$ one
obtains the quasi-de Sitter solution $H(t)=H_0-H_i t$, which is
indeed an inflationary solution known to provide viable
cosmological solutions in $f(R)$ gravity. However, our strategy is
to assume that $m=0$ without solving explicitly the differential
equation (\ref{parameterm}). As we already mentioned, there are
many cosmologies that may yield $m=0$ apart from the quasi-de
Sitter one, for example the symmetric bounce $a(t)=e^{-\Lambda
t}$, in which case $H(t)=-2\Lambda t$, and the bouncing point
might be the time instance $t=0$. In that case, the solution for
the Hubble rate is of the form $H(t)=H_0-H_i t$, but with $H_0$
being set equal to zero. So the case $m=0$ describes a class of
cosmological models, not a specific one. Now by setting $m=0$, and
by studying the behavior of the dynamical system, one may reveal
what is the nature of the cosmological solution, by finding the
fixed points and evaluating for these the EoS parameter $w_{eff}$
given in Eq. (\ref{eos1}). This is exactly what we did in Ref.
\cite{Odintsov:2017tbc}, where we demonstrated that the case $m=0$
in the flat space $f(R)$ gravity dynamical system, leads to fixed
points which have $w_{eff}=-1$. Thus the dynamical system
indicated directly what is the nature of the subclass of
cosmological solutions corresponding to $m=0$. The method is
powerful and formal since we do not fix the form of the Hubble
rate, by specifying the initial conditions for example (that is we
do not fix $H_0$ and $H_i$ in the solution $H(t)=H_0-H_it$), and
the dynamical system analysis reveals the nature of the resulting
attractor solution, by yielding $w_{eff}=-1$, which is an exact de
Sitter final attractor. Also it is important to note that the
system may start from a quasi-de Sitter solution, and it results
to an exact de Sitter solution, as was demonstrated in Ref.
\cite{Odintsov:2017tbc}. Thus the method of dynamical systems
reveals the actual evolution of the cosmological solution without
making any specific assumptions on the initial conditions chosen
for  the Hubble rate, apart from the obvious choice of the
subclass of solutions corresponding to $m=0$. In conclusion, one
does not specify the exact value of the Hubble rate by setting
$m=0$, but a subclass of cosmological solutions is studied. The
dynamical system analysis reveals the exact evolution in terms of
the fixed points and the corresponding phase space trajectories.

As was shown in Ref. \cite{Odintsov:2017tbc}, for the case $m=0$,
the dynamical system (\ref{dynamicalsystemmain}) has the following
two fixed points,
\begin{equation}\label{fixedpointdesitter}
\phi_*^1=(x_1,x_2,x_3)=(-1,0,2),\,\,\,\phi_*^2=(x_1,x_2,x_3)=(0,-1,2)\,
,
\end{equation}
and as it was shown numerically, for various initial conditions,
the fixed point $\phi_*^1=(x_1,x_2,x_3)=(-1,0,2)$ is stable and
the fixed point $\phi_*^2(x_1,x_2,x_3)=(0,-1,2)$ is unstable.
Also, by substituting the value of $x_3$ corresponding to both
fixed points (\ref{fixedpointdesitter}), namely  $x_3=2$, we
obtain $w_{eff}=-1$, so both the fixed points $\phi_*^1$ and
$\phi_*^2$ are de Sitter fixed points, or equivalently, de Sitter
attractors. Moreover, as was shown in Ref.
\cite{Odintsov:2017tbc}, the vacuum $f(R)$ gravity which leads to
the fixed point $\phi_*^1$ has the following functional form,
\begin{equation}\label{solapprox1}
f(R)\simeq \Lambda_1-24 \Lambda_2e^{-\frac{R}{24H_i}}\, ,
\end{equation}
while the functional form of the $f(R)$ gravity which leads to the
fixed point $\phi_*^2=(0,-1,2)$ at leading order, is the
following,
\begin{equation}\label{approximatersquare}
f(R)\simeq \alpha R^2\, .
\end{equation}
Finally, the variables $x_1$, $x_2$ and $x_3$
(\ref{variablesslowdown}), satisfy the Friedmann constraint,
\begin{equation}\label{friedmannconstraint}
x_1+x_2+x_3=1\, ,
\end{equation}
which is also satisfied by the fixed points. Having discussed the
flat vacuum $f(R)$ gravity phase space structure, in the next
section we present in detail the effects of spatial curvature on
the phase space structure of vacuum $f(R)$ gravity.

\section{Autonomous Vacuum $f(R)$ Gravity Dynamical System in Spatially Curved Spacetime }

In this section we shall investigate how it is possible to
construct an autonomous dynamical system for a vacuum $f(R)$
gravity theory, for a spatially non-flat FRW metric. The action of
the theory and the equations of motion for general metric remain
the same as in Eqs. (\ref{action}) and
(\ref{modifiedeinsteineqns}) respectively. So let us assume that
the metric has the general non-spatially flat FRW form,
\begin{equation}\label{frwnonflat}
ds^2 = - dt^2 + a(t)^2
\left(\frac{dr^2}{1-Kr^2}+r^2d\theta^2+r^2\sin^2 \theta
d\varphi^2\right)\, ,
\end{equation}
with $a(t)$ being the scale factor and $K$ is the curvature of the
three dimensional spacelike hypersurface $t=$const, which can be
$K=0,\pm 1$. In this case, the Ricci scalar has the following
form,
\begin{equation}\label{ricciscalaranalyticcurved}
R=6\dot{H}+12H^2 +\frac{6K}{a^2}\, ,
\end{equation}
where we have used a physical units system in which $\kappa^2=8\pi
G=1$, and we adopt this notation for simplicity in the following.
For the metric (\ref{frwnonflat}), the equations of motion
(\ref{modifiedeinsteineqns}) become,
\begin{align}\label{JGRG15curvedspace}
& 3F H^2+\frac{3KF}{a^2}=-\frac{f-F\,R}{2}-3H\dot{F}\, , \\ \notag
& \ddot{F}=H\dot{F}-2F\dot{H}+\frac{2KF}{a^2}\, ,
\end{align}
where again $F(R)=\frac{\partial f}{\partial R}$,
$F'(R)=\frac{\partial F}{\partial R}$, and
$F''(R)=\frac{\partial^2 F}{\partial R^2}$. Two useful relations
for the extraction of the autonomous dynamical system
corresponding to the cosmological equations
(\ref{JGRG15curvedspace}) are the following,
\begin{align}\label{usefulrelations}
& \dot{H}=\frac{R}{6}-2H^2-\frac{K}{a^2},
\\ \notag & \dot{R}=24 H\dot{H}+6\ddot{H}-\frac{12KH}{a^2}\, ,
\end{align}
which we shall extensively make use of in the following. In order
to extract an autonomous dynamical system from Eqs.
(\ref{JGRG15curvedspace}), it is compelling to introduce the
following variables (recall we have set $\kappa^2=1$),
\begin{equation}\label{variablesslowdowncurved}
x_1=-\frac{\dot{F}(R)}{F(R)H},\,\,\,x_2=-\frac{f(R)}{6F(R)H^2},\,\,\,x_3=
\frac{R}{6H^2},\,\,\,x_4=-\frac{K}{a^2H^2}\, ,
\end{equation}
so by comparing Eqs. (\ref{variablesslowdown}) and
(\ref{variablesslowdowncurved}) it can be seen that due to the
non-zero curvature effects, it is required to use an additional
variable in order to extract an autonomous dynamical system from
the $f(R)$ cosmological equations. Again, by using the
$e$-foldings number as a dynamical variable, by using the new
variables (\ref{variablesslowdowncurved}) and the cosmological
equations (\ref{JGRG15curvedspace}), we obtain the following
autonomous dynamical system for the vacuum $f(R)$ gravity in a
spatially non-flat FRW background,
\begin{align}\label{dynamicalsystemmaincurved}
&
\frac{\mathrm{d}x_1}{\mathrm{d}N}=-4+3x_1+2x_3-x_1x_3+x_1^2+4x_4-x_1x_4\,
,
\\ \notag &
\frac{\mathrm{d}x_2}{\mathrm{d}N}=8+m-4x_3+x_2x_1-2x_2x_3+4x_2-6x_4-2x_2x_4 \, ,\\
\notag & \frac{\mathrm{d}x_3}{\mathrm{d}N}=-8-m+8x_3-2x_3^2+6x_4-2x_3x_4 \, ,\\
\notag & \frac{\mathrm{d}x_4}{\mathrm{d}N}=-2 x_3 x_4-2 x_4^2+4
x_4-2 x_4\, ,
\end{align}
where we used Eqs. (\ref{JGRG15curvedspace}),
(\ref{usefulrelations}), (\ref{variablesslowdowncurved}) and
(\ref{specialderivative}) for the extraction of the dynamical
system, and in this case too, the parameter $m$ is given in Eq.
(\ref{parameterm}). As in the flat FRW metric case, the only
$N$-dependence of the dynamical system
(\ref{dynamicalsystemmaincurved}) is contained in the parameter
$m$, so if we assume that this is constant, then the dynamical
system is rendered autonomous. The nature of the possible fixed
points of the dynamical system (\ref{dynamicalsystemmaincurved})
will be determined by the total EoS parameter $w_{eff}$, which in
the non-flat FRW dynamical system case has the following form,
\begin{equation}\label{eos1curved}
w_{eff}=-1-\frac{2}{3} (x_3-2+x_4)\, ,
\end{equation}
a relation which easily follows if Eqs. (\ref{weffoneeqn}),
(\ref{usefulrelations}) and (\ref{variablesslowdowncurved}) are
combined. Finally, the Friedmann constraint in this case reads,
\begin{equation}\label{friedmannconstraintcurved}
x_1+x_2+x_3+x_4=1\, .
\end{equation}
In the following sections we shall investigate the structure of
the phase space for the non-flat FRW vacuum $f(R)$ gravity, for
various constant values of the parameter $m$, emphasizing on the
cases $m=0$, $m=-\frac{9}{2}$ and $m=-8$, which describe de
Sitter, matter  and radiation domination fixed points.

\subsection{Study de Sitter Attractors of Spatially Curved $f(R)$ Gravity Phase Space}

Let us focus our study in the case  $m=0$, which as we saw in the
flat vacuum $f(R)$ gravity corresponds to de Sitter fixed points.
So in this section we will address the following questions, are
the fixed points of the non-flat vacuum $f(R)$ gravity dynamical
system (\ref{dynamicalsystemmaincurved}) de Sitter vacua, and if
yes, are these stable, or the presence of the curvature terms
destabilizes the phase space? For $m=0$, the dynamical system
(\ref{dynamicalsystemmaincurved}) is rendered autonomous, so the
study of the behavior of the trajectories in the phase space can
be studied in a easy and formal way, by examining the linearized
system. Let us recall the formalism of the Hartman-Grobman
theorem, which will be useful in the following sections. Consider
a dynamical system of the form,
\begin{equation}\label{ds1}
\frac{\mathrm{d}\Phi}{\mathrm{d}t}=g(\Phi (t))\, ,
\end{equation}
with $g(\Phi (t))$  being a locally Lipschitz continuous map of
the following form $g:R^n\rightarrow R^n$. Let the fixed points of
the autonomous dynamical system (\ref{ds1}) be $\phi_*$. The
Jacobian matrix of the linearized dynamical system near a fixed
point, denoted as $\mathcal{J}(g)$, is equal to,
\begin{equation}\label{jaconiab}
\mathcal{J}=\sum_i\sum_j\Big{[}\frac{\mathrm{\partial
g_i}}{\partial x_j}\Big{]}\, .
\end{equation}
By calculating the Jacobian at the fixed points, this will
eventually reveal whether a fixed point is stable or not, only in
the case that the fixed point is hyperbolic. A hyperbolic fixed
point corresponds to the case that the eigenvalues of the Jacobian
matrix at the fixed point, namely $e_i$, have non-zero real parts,
that is, $\mathrm{Re}(e_i)\neq 0$. Then the Hartman-Grobman
ensures that the linearized dynamical system,
\begin{equation}\label{loveisalie}
\frac{\mathrm{d}\Phi}{\mathrm{d}t}=\mathcal{J}(g)(\Phi)\Big{|}_{\Phi=\phi_*}
(\Phi-\phi_*)\, ,
\end{equation}
is locally homeomorphic to the initial dynamical system
(\ref{ds1}), at the vicinity of the fixed points $\phi_*$, hence
by studying the linearized dynamical system is enough in order to
understand the behavior of the trajectories near the fixed points,
and therefore decide about their stability. If the eigenvalues of
the Jacobian satisfy $\mathrm{Re}\left(\sigma
(\mathcal{J}(g))\right)<0$, then the fixed point is stable, and in
all other cases, the fixed point is unstable.

\begin{figure}[h]
\centering
\includegraphics[width=18pc]{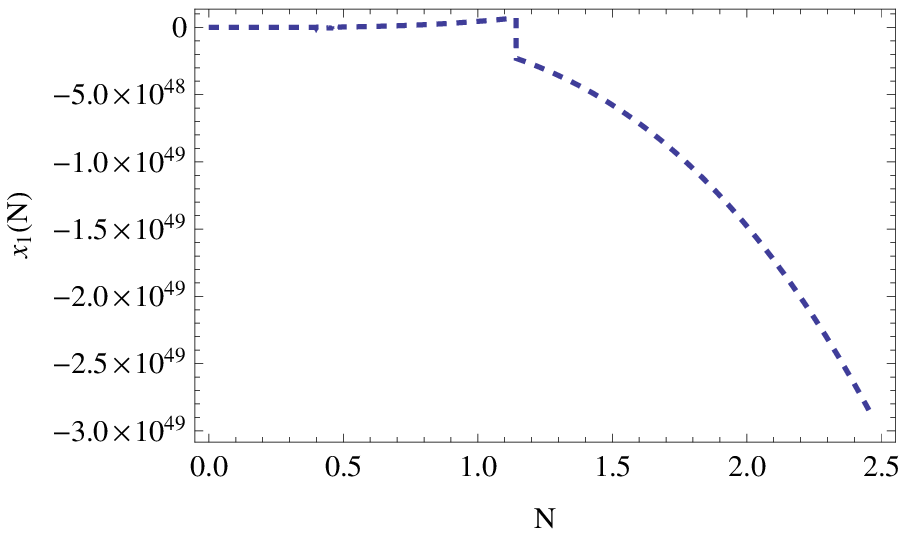}
\includegraphics[width=18pc]{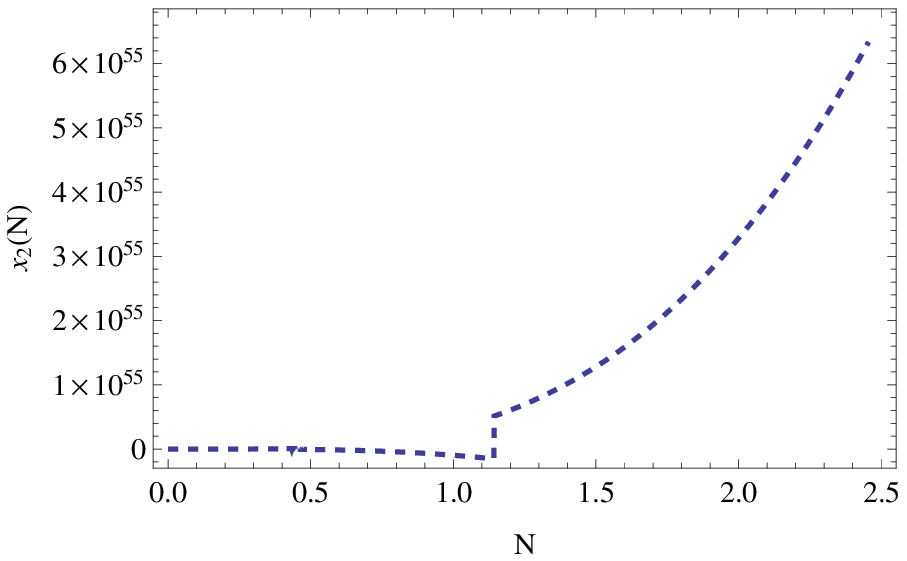}
\includegraphics[width=18pc]{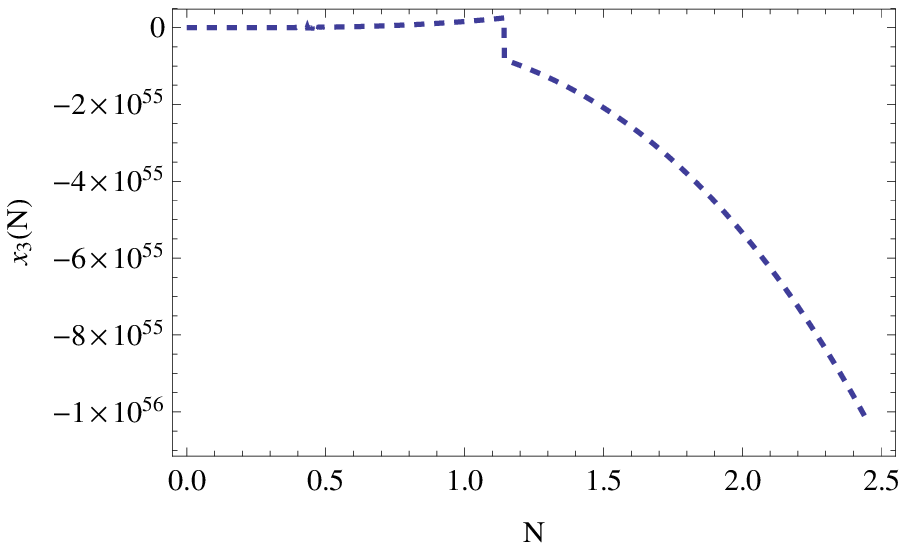}
\includegraphics[width=18pc]{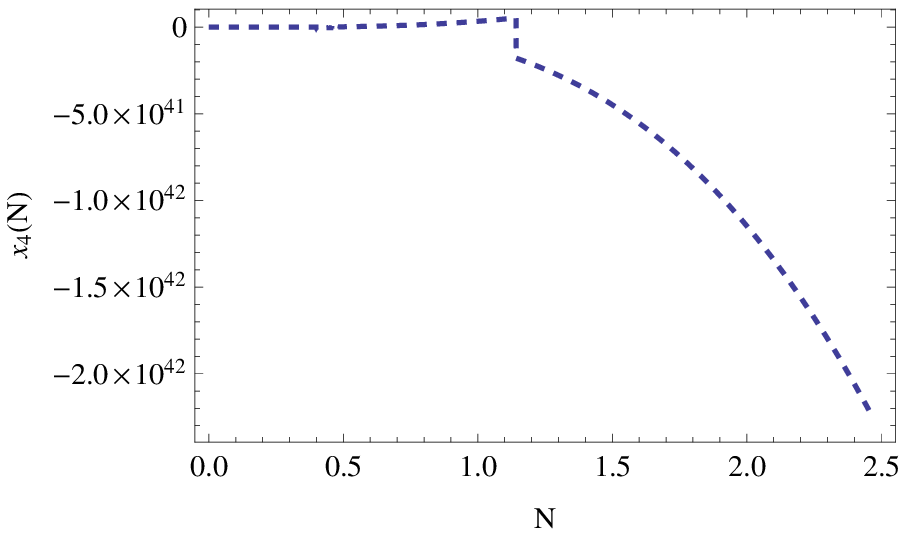}
\caption{{\it{The behavior of $x_1(N)$, $x_2(N)$, $x_3(N)$ and
$x_4(N)$ as functions of the $e$-foldings, for the non-flat FRW
autonomous vacuum $f(R)$ gravity dynamical system in the case
$m=0$.}}} \label{plot1}
\end{figure}

\begin{figure}[h]
\centering
\includegraphics[width=18pc]{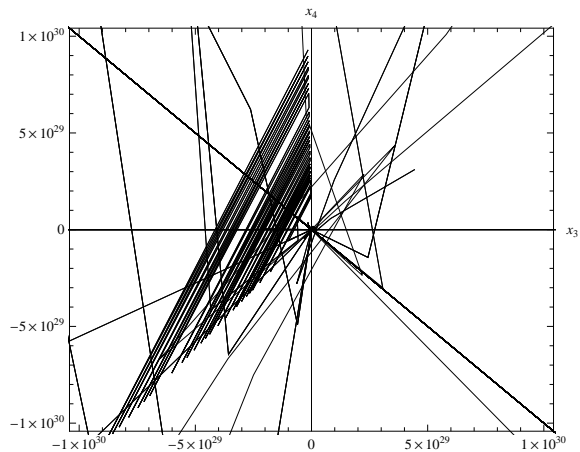}
\caption{{\it{The trajectories in the $x_3-x_4$ plane for various
initial conditions, for the non-flat FRW autonomous vacuum $f(R)$
gravity dynamical system in the case $m=0$.}}} \label{plot2}
\end{figure}

Let us now find the fixed points of the dynamical system
(\ref{dynamicalsystemmaincurved}) in the case $m=0$, and let the
functions $f_i$ for general $m$, be defined as follows,
\begin{align}\label{functionsfi}
& f_1(x_1,x_2,x_3,x_4)= -4+3x_1+2x_3-x_1x_3+x_1^2+4x_4-x_1x_4,
\\ \notag & f_2(x_1,x_2,x_3,x_4)=8+m-4x_3+x_2x_1-2x_2x_3+4x_2-6x_4-2x_2x_4 ,
\\ \notag
& f_3(x_1,x_2,x_3,x_4)=-8-m+8x_3-2x_3^2+6x_4-2x_3x_4 ,
\\ \notag & f_4(x_1,x_2,x_3,x_4)=-2 x_3 x_4-2 x_4^2+4
x_4-2 x_4 ,
\end{align}
and the corresponding Jacobian matrix,
\begin{equation}\label{jaconiab}
\mathcal{J}=\sum_{i=1}^4\sum_{j=1}^4\Big{[}\frac{\mathrm{\partial
f_i}}{\partial x_j}\Big{]}\, ,
\end{equation}
is equal to,
\begin{equation}\label{jacobiangeneral}
\mathcal{J}=\left(
\begin{array}{cccc}
 2 x_1-x_3-x_4+3 & 0 & 2-x_1 & 4-x_1 \\
 x_2 & x_1-2 x_3-2 x_4+4 & -2 x_2-4 & -2 x_2-6 \\
 0 & 0 & -4 x_3-2 x_4+8 & 6-2 x_3 \\
 0 & 0 & -2 x_4 & -2 x_3-4 x_4+2 \\
\end{array}
\right)\, .
\end{equation}
Focusing now in the case $m=0$, by solving the system of equations
$f_i=0$, $i=1,..,4$, we obtain the following fixed points with
physical significance,
\begin{equation}\label{fixedpointsdesittercurvedaux}
\phi_*^1=(x_1,x_2,x_3,x_4)=(-1,0,2,0),\,\,\,\phi_*^2=(x_1,x_2,x_3,x_4)=(0,x_2,2,0)\,
.
\end{equation}
By looking at Eqs. (\ref{fixedpointsdesittercurvedaux}) and
(\ref{fixedpointdesitter}), the similarity is obvious, since the
only difference is the existence of the condition $x_4=0$ in the
new fixed points of the phase space. Also for the fixed point
$\phi_*^2$, the values of $x_2$ are free to be chosen, but in
order to have consistency with the Friedmann constraint
(\ref{friedmannconstraintcurved}), we must require $x_2=-1$ for
the fixed point $\phi_*^2$, therefore the resulting fixed points
for the case $m=0$ are,
\begin{equation}\label{fixedpointsdesittercurved}
\phi_*^1=(x_1,x_2,x_3,x_4)=(-1,0,2,0),\,\,\,\phi_*^2=(x_1,x_2,x_3,x_4)=(0,-1,2,0)\,
.
\end{equation}
Clearly, for both the fixed points, the EoS parameter
(\ref{eos1curved}) is equal to $w_{eff}=-1$ when evaluated at the
fixed points (\ref{fixedpointsdesittercurved}), so both the fixed
points (\ref{fixedpointsdesittercurved}) are de Sitter fixed
points. As we saw in the previous section, for the flat FRW vacuum
$f(R)$ gravity case, the fixed point $\phi_*^1$ was found to be
stable, while the fixed point $\phi_*^2$ was unstable. Let us now
see what new features does the variable $x_4$ introduces in the
phase space. In order to see this, let us calculate the
eigenvalues of the Jacobian at the fixed points, so we start off
with $\phi_*^1$, so the Jacobian in this case is,
\begin{equation}\label{jacobianforfixedpointf1}
\mathcal{J}=\left(
\begin{array}{cccc}
 -1 & 0 & 3 & 5 \\
 0 & -1 & -4 & -6 \\
 0 & 0 & 0 & 2 \\
 0 & 0 & 0 & -2 \\
\end{array}
\right)\, ,
\end{equation}
and the corresponding eigenvalues are $(-2,-1, -1, 0)$, while for
the fixed point $\phi_*^2$, the Jacobian is,
\begin{equation}\label{jacobianfixedpoint2}
\mathcal{J}=\left(
\begin{array}{cccc}
 1 & 0 & 2 & 4 \\
 -1 & 0 & -2 & -4 \\
 0 & 0 & 0 & 2 \\
 0 & 0 & 0 & -2 \\
\end{array}
\right)\, ,
\end{equation}
with eigenvalues $(-2, 1, 0, 0)$. Therefore, by looking the
eigenvalues it is obvious that both fixed point are not
hyperbolic, and in effect, numerical analysis is needed to reveal
the stability of the fixed points. After a thorough investigation
of the trajectories in the phase space spanned by the variables
$x_1$, $x_2$, $x_3$ and $x_4$, for various initial conditions, the
resulting picture is that the trajectories of all the variables
blow-up after a few $e$-foldings, so in effect, the two fixed
points (\ref{fixedpointsdesittercurved}) are strongly unstable.
This can also be seen in Fig. \ref{plot1}, where we plotted the
behavior of $x_1(N)$, $x_2(N)$, $x_3(N)$ and $x_4(N)$ as functions
of the $e$-foldings.
As it can be seen in Fig. \ref{plot1}, the trajectories blow-up
even after a few $e$-foldings, and this occurs for various initial
conditions, for both positive and negative curvatures, even for
small initial values. Hence, we can conclude that the effect of
spatial curvature in the vacuum $f(R)$ gravity phase space, is
that it completely destabilizes the stable de Sitter fixed points
of the flat case. The result is model independent and covers all
possible cases of $f(R)$ gravity that may lead to de Sitter vacua.
This can also be seen for various initial conditions in Fig.
\ref{plot2}, where we plotted the trajectories in the $x_3-x_4$
plane. As it can be seen, there exist multiple infinite
trajectories, a fact that depicts the strong instabilities that
the curvature introduces in the phase space.
Before closing it is worth noting that in this case too, the
functional form of the $f(R)$ gravity leading to the fixed points
$\phi_*^1$ and $\phi_*^2$ (\ref{fixedpointsdesittercurved}), are
given in Eqs. (\ref{solapprox1}) and (\ref{approximatersquare})
respectively. More importantly, both the fixed points
(\ref{fixedpointsdesittercurved}) have $x_4=0$. Thus the dynamical
system has essentially the same fixed points with the flat case
$f(R)$ gravity, however the effect of the curvature is non-trivial
in the dynamical system, since it destabilizes the fixed points.
The dynamical system of the non-flat case obviously has a rich
underlying structure, possibly in the form of some unstable
manifolds, the study of which is beyond the scopes of this
article.

Before closing we need to discuss an important issue having to do
with the dimensions of the dynamical system
(\ref{dynamicalsystemmaincurved}) and the Friedman constraint
(\ref{friedmannconstraintcurved}). Basically one could reduce the
dimensions of the dynamical system to three instead of four by
taking into account the Friedmann constraint, however we kept all
the variables in order to have a concrete idea on how these behave
as functions of the $e$-foldings number, and see how the
instability occurs in the system. In the case at hand, the
presence of the $x_4$ terms make the system strongly unstable, so
there surely is some unstable submanifold related to $x_4$. We
hope to address these mathematical issues in a focused future
work.

\subsection{Study of Matter and Radiation Domination Era Attractors of Spatially Curved $f(R)$ Gravity Phase Space}

Now let us turn our focus on other cosmological solutions, so let
us discuss the case $m=-\frac{9}{2}$, which as was shown in Ref.
\cite{Odintsov:2017tbc}, corresponds to matter domination fixed
points of the autonomous dynamical system, with $w_{eff}=0$. The
fixed points of the corresponding flat autonomous dynamical system
(\ref{dynamicalsystemmain}) with $m=-\frac{9}{2}$, are the
following \cite{Odintsov:2017tbc},
\begin{equation}\label{fixedpointdesitter1}
\phi_*^1=(\frac{1}{4} \left(-5-\sqrt{73}\right),\frac{1}{4}
\left(7+\sqrt{73}\right),\frac{1}{2}),\,\,\,\phi_*^2=(\frac{1}{4}
\left(\sqrt{73}-5\right),\frac{1}{4}
\left(7-\sqrt{73}\right),\frac{1}{2})\, .
\end{equation}
The eigenvalues of the Jacobian matrix $\mathcal{J}$ for the first
fixed point $\phi_*^1$ are $(6,-4.272,-0.386001)$, while for the
fixed point $\phi_*^2$ these are $(6,4.272,3.886)$. In effect,
both equilibria are hyperbolic and unstable. Also, since
$x_3=\frac{1}{2}$, for both the fixed points
(\ref{fixedpointdesitter1}), in both cases the EoS parameter is
$w_{eff}=0$, and this justifies why these fixed points are called
matter domination fixed points.

\begin{figure}[h]
\centering
\includegraphics[width=18pc]{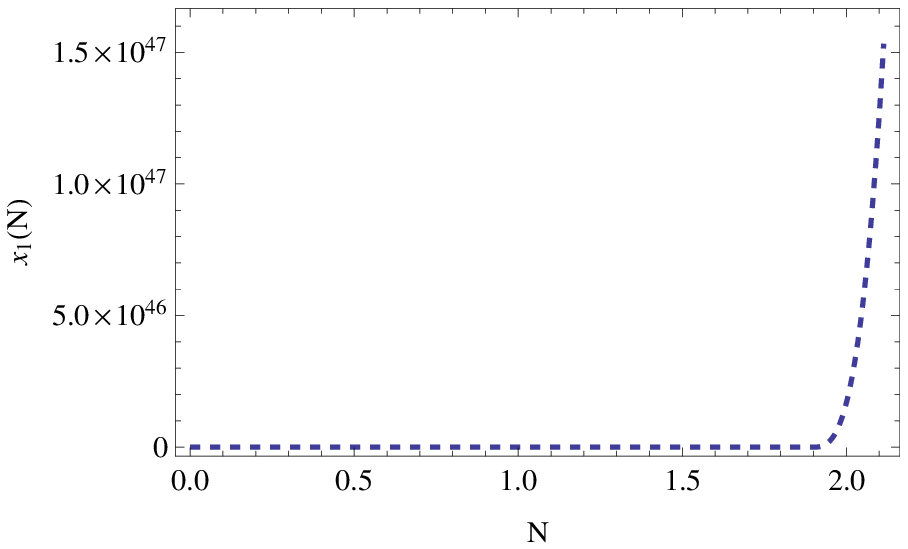}
\includegraphics[width=18pc]{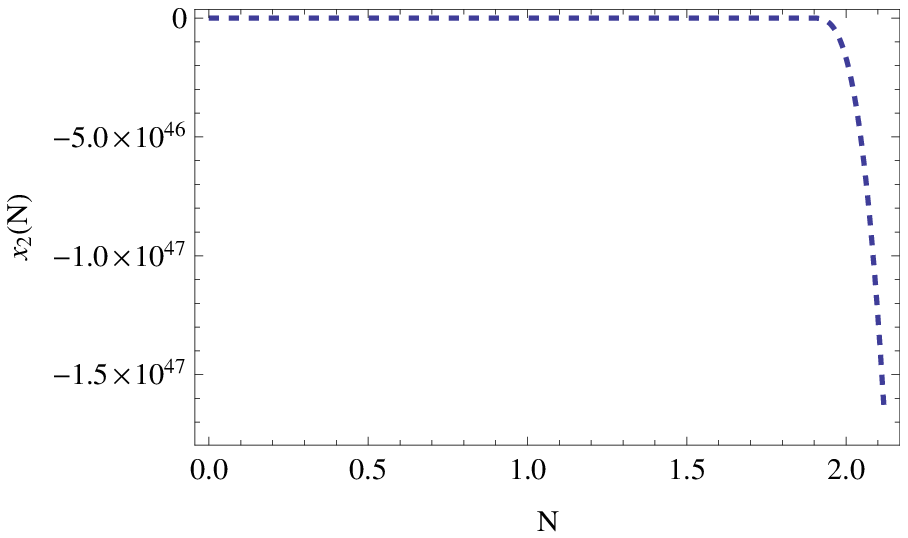}
\includegraphics[width=18pc]{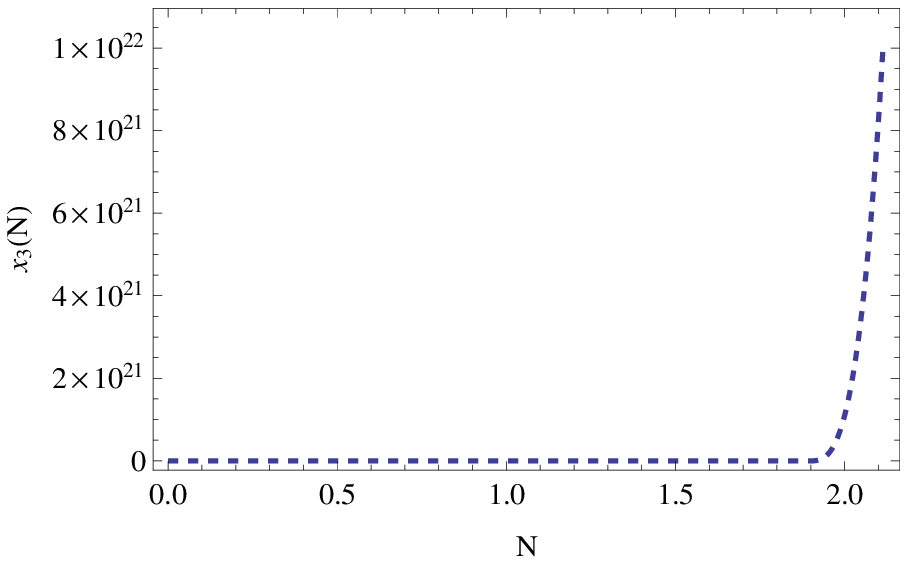}
\includegraphics[width=18pc]{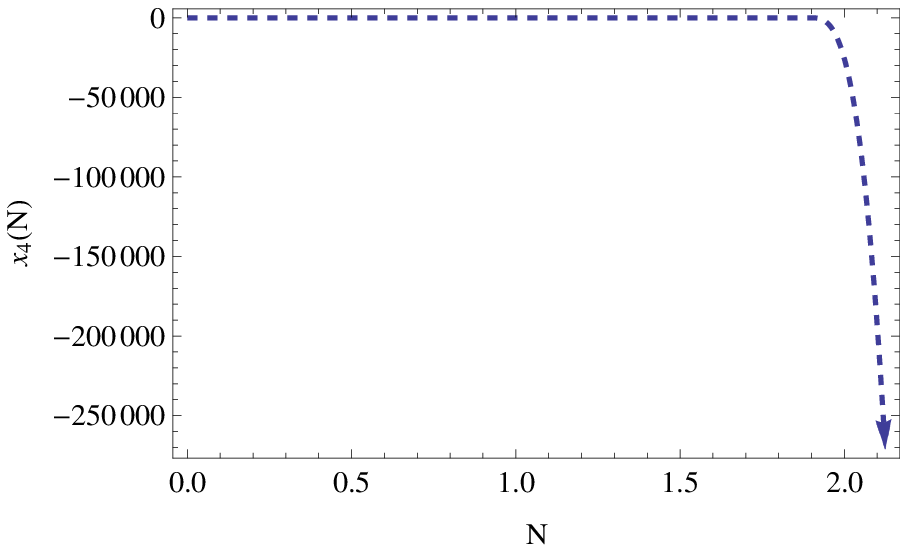}
\caption{{\it{The behavior of $x_1(N)$, $x_2(N)$, $x_3(N)$ and
$x_4(N)$ as functions of the $e$-foldings, for the non-flat FRW
autonomous vacuum $f(R)$ gravity dynamical system in the case
$m=-\frac{9}{2}$.}}} \label{plot3}
\end{figure}

Having discussed the flat autonomous $f(R)$ gravity dynamical
system for the case $m=-\frac{9}{2}$, let us now see the effects
of the spatial curvature on the corresponding autonomous dynamical
system (\ref{dynamicalsystemmaincurved}). For $m=-\frac{9}{2}$ the
physically acceptable fixed points of the dynamical system are the
following,
\begin{align}\label{fixedpointcurvedmatt}
& \phi_*^1=(x_1,x_2,x_3,x_4)=(\frac{1}{4}
\left(-5-\sqrt{73}\right),\frac{1}{2} \left(\frac{1}{2}
\left(5+\sqrt{73}\right)+1\right),\frac{1}{2},0), \\ \notag &
\phi_*^2=(x_1,x_2,x_3,x_4)=(\frac{1}{4}
\left(\sqrt{73}-5\right),\frac{1}{2} \left(\frac{1}{2}
\left(5-\sqrt{73}\right)+1\right),\frac{1}{2},0)\, .
\end{align}
As it can be easily checked, for both the fixed points $\phi_*^1$
and $\phi_*^2$, the EoS parameter (\ref{eos1curved}) becomes
$w_{eff}=0$, so both the fixed points (\ref{fixedpointcurvedmatt})
are matter domination fixed points. Also the Friedmann constraint
(\ref{friedmannconstraintcurved}) is also satisfied for both the
fixed points. For the fixed point $\phi_*^1$ the Jacobian matrix
(\ref{jacobiangeneral}) becomes,
\begin{equation}\label{jacobianmatter1}
\mathcal{J}=\left(
\begin{array}{cccc}
 -\frac{\sqrt{73}}{2} & 0 & \frac{1}{4} \left(13+\sqrt{73}\right) & \frac{1}{4} \left(21+\sqrt{73}\right) \\
 \frac{1}{4} \left(7+\sqrt{73}\right) & \frac{1}{4} \left(7-\sqrt{73}\right) & \frac{1}{2} \left(-15-\sqrt{73}\right) & \frac{1}{2} \left(-19-\sqrt{73}\right) \\
 0 & 0 & 6 & 5 \\
 0 & 0 & 0 & 1 \\
\end{array}
\right)\, ,
\end{equation}
and it's eigenvalues are $(6,-4.272,1,-0.386001)$. In effect, the
fixed point $\phi_*^1$ is a hyperbolic and unstable fixed point.
Accordingly, for the fixed point $\phi_*^2$, the Jacobian matrix
is,
\begin{equation}\label{jacobianmatter2}
\mathcal{J}=\left(
\begin{array}{cccc}
 \frac{\sqrt{73}}{2} & 0 & \frac{1}{4} \left(13-\sqrt{73}\right) & \frac{1}{4} \left(21-\sqrt{73}\right) \\
 \frac{1}{4} \left(7-\sqrt{73}\right) & \frac{1}{4} \left(7+\sqrt{73}\right) & \frac{1}{2} \left(-15+\sqrt{73}\right) & \frac{1}{2} \left(-19+\sqrt{73}\right) \\
 0 & 0 & 6 & 5 \\
 0 & 0 & 0 & 1 \\
\end{array}
\right)\, ,
\end{equation}
and the corresponding eigenvalues are $(6, 4.272, 3.886, 1)$, so
the fixed point $\phi_*^2$ is also hyperbolic and unstable. The
instability of the phase space can also be seen in Fig.
\ref{plot3}, where we plot the functional dependence of the
variables $x_i(N)$, $i=1,..,4$, by solving numerically the
dynamical system equations.

Similar considerations can be made for radiation domination fixed
points. By choosing $m=-8$, the fixed points of the dynamical
system (\ref{dynamicalsystemmaincurved}) are the following,
\begin{align}\label{fixedpointcurvedmatt1}
& \phi_*^1=(x_1,x_2,x_3,x_4)=(-4,5,0,0), \\ \notag &
\phi_*^2=(x_1,x_2,x_3,x_4)=(1,0,0,0)\, .
\end{align}
For both the fixed points $\phi_*^1$ and $\phi_*^2$, the EoS
parameter (\ref{eos1curved}) becomes $w_{eff}=\frac{1}{3}$, so
both the fixed points (\ref{fixedpointcurvedmatt1}) are radiation
domination fixed points. For the fixed point $\phi_*^1$ the
Jacobian matrix (\ref{jacobiangeneral}) is equal to,
\begin{equation}\label{jacobianmatter11}
\mathcal{J}=\left(
\begin{array}{cccc}
 -5 & 0 & 6 & 8 \\
 5 & 0 & -14 & -16 \\
 0 & 0 & 8 & 6 \\
 0 & 0 & 0 & 2 \\
\end{array}
\right)\, ,
\end{equation}
the eigenvalues of which are $(8, -5, 2, 0)$. Also for the fixed
point $\phi_*^2$, the Jacobian matrix is,
\begin{equation}\label{jacobianmatter22}
\mathcal{J}=\left(
\begin{array}{cccc}
 5 & 0 & 1 & 3 \\
 0 & 5 & -4 & -6 \\
 0 & 0 & 8 & 6 \\
 0 & 0 & 0 & 2 \\
\end{array}
\right)\, ,
\end{equation}
and the eigenvalues are $(8, 5, 5, 2)$, so only the fixed point
$\phi_*^2$ is hyperbolic and it is clearly unstable. Finally, a
numerical analysis of the phase space indicates that strong
instabilities occur in the phase space.

In conclusion, apart from the de Sitter fixed points studied in
the previous section, also the matter and radiation domination
fixed points are unstable equilibria of the autonomous curved FRW
vacuum $f(R)$ gravity dynamical system. Hence, this result
supports our claim that the effect of curvature is to destabilize
the phase space structure of vacuum $f(R)$ gravity.

Qualitatively, our results clearly indicate that the stable de
Sitter attractors of the flat spacetime vacuum $f(R)$ gravity are
destabilized strongly even in the presence of a small non-zero
curvature in the metric. This result holds true even for positive
(elliptic spatial sections) and negative curvatures (hyperbolic
spatial curvatures). Hence due to the unstable phase space
structure, we have strong hints to claim that inflation in the
non-flat vacuum $f(R)$ gravity theories is problematic or at least
is not so universal, compared to the flat case one, since there is
no stable de Sitter attractor in the theory. The existence of
stable inflationary attractors in a gravitational theory is of
fundamental importance, since even the slow-roll expansion assumes
a stable attractor that eventually will attract all the phase
space trajectories on it. An insightful work analyzing the
slow-roll expansion in view of the existence of inflationary
attractors is \cite{Liddle:1994dx}.

\section{Conclusions}

In this paper we studied the effects of spatial curvature in the
autonomous dynamical system of vacuum $f(R)$ gravity. After
appropriately choosing the variables, we constructed a dynamical
system which can be rendered autonomous in the case that the
parameter $m=-\frac{\ddot{H}}{H^3}$ takes constant values. We
focused on three cases of interest, namely for $m=0$,
$m=-\frac{9}{2}$ and $m=-8$. The case $m=0$ corresponds to de
Sitter vacua, since the fixed points have a EoS parameter equal to
$w_{eff}=-1$, which is characteristic of the de Sitter evolution.
As we demonstrated, the presence of spatial curvature destabilizes
the flat space phase space, which was shown in a previous work to
be stable. The result is robust towards the choice of different
initial conditions. The absence of stable inflationary attractors
indicates that inflation in non-flat vacuum $f(R)$ gravity is
problematic and not so universal compared with the flat vacuum
$f(R)$ gravity case. The same instability behavior occurs for the
case $m=-\frac{9}{2}$, which corresponds to matter domination
related fixed points, and also the same behavior applies in the
case $m=-8$, which describes radiation domination equilibria. As
we demonstrated, all the matter and radiation domination fixed
points are unstable hyperbolic equilibria. However, the flat case
autonomous dynamical system for $f(R)$ gravity, in the cases of
matter and radiation domination equilibria, was also unstable, so
the most interesting case of this study was the de Sitter case,
and particularly one needs to understand why the non-flat
autonomous dynamical system becomes destabilized. It is clear that
the new variable $x_4$, which was absent in the flat case, brings
some instability in the dynamical system. So it is certain that
the phase space of the non-flat de Sitter case has a rich
underlying mathematical structure. It is highly possible that
non-stable manifolds are formed, strongly related with the $x_4$
subspace. The study of these unstable manifolds is mathematically
non-trivial because the dynamical system is highly non-linear, but
the complete investigation of the above issues will be given
elsewhere. Finally, it would be interesting to extend the current
study by taking into account dark matter and baryonic matter.
Eventually, in this case the dynamical system would become much
more complicated and only numerical investigation would be the
only approach for studying the dynamical evolution of the
trajectories in the phase space.

\section*{Acknowledgments}

This work is supported by MINECO (Spain), FIS2016-76363-P (S.D.O),
by project 2017 SGR247 (AGAUR, Catalonia) (S.D. Odintsov).

\end{document}